\documentclass[12pt]{iopart}

\usepackage{multirow,graphicx,url,braket,bm}
\usepackage{dcolumn}
\usepackage{hyperref}%
\begin{document}

\title{Weakening of $N=28$ shell gap and the nature of $0_2^+$ states}

\author{Bhoomika Maheshwari$\star$}
\address{Department of Physics, 
Faculty of Science, University of Zagreb, 
Zagreb, HR-10000, Croatia}
\ead{bhoomika.physics@gmail.com \\ $\star:$Corresponding author}

\author{Kosuke Nomura}
\address{Department of Physics, 
Hokkaido University, Sapporo 060-0810, Japan}
\address{Nuclear Reaction Data Center, 
Hokkaido University, Sapporo 060-0810, Japan}
\ead{nomura@sci.hokudai.ac.jp}

\begin{abstract}
The work reports a novel application of 
the interacting boson model 
in light-mass region around $^{48}$Ca, 
that takes into account intruder states and 
configuration mixing. 
The model is shown to provide a reasonable 
description of the observed 
low-lying yrast and yrare states of 
the $N=28$ even-even isotones from Si to Fe, 
and even-even Ca isotopes.  
The nature of $0_2^+$ states is addressed 
in terms of the competition between 
spherical and deformed intruder configurations, 
and the rigidity of the $N=28$ shell gap is tested, particularly for $^{44}$S. 
The $0_2^+$ isomer in $^{44}$S is shown 
to arise from a weak mixing between 
the two configurations and called as shape isomer. 
The unresolved low-lying spectra in $^{46}$Ar 
is also approached using the core-excitations 
across $N=28$ shell. 
The SU(3) structure in $^{42}$Si is supported 
by the present calculation. 
The nearly spherical nature of $^{50}$Ti, 
$^{52}$Cr, $^{54}$Fe and $^{42,44,46}$Ca 
isotopes is found, 
while the core excitations are found 
to be essential for the description of yrare states. 
Shell model calculation is also performed for comparison.  
\end{abstract}

%
%
%
%
%

\section{Introduction}

The atomic nuclei, as well as atoms, 
exhibit a few regular 
shell gaps, which give rise to extra stability. 
The concept of shell gaps for protons and neutrons 
at magic numbers stands as the backbone 
of modeling nuclear structure. 
The magic numbers emerge from the strong 
spin-orbit interaction, which is responsible 
for the extra binding of nucleons in orbitals 
having angular momentum aligned with 
their intrinsic spin \cite{mayer1955}. 
These gaps are usually large enough to prevent 
excitations, but in lighter nuclei, 
nucleonic excitations across the magic-core 
are more likely to happen due to smaller 
shell gaps between the low-$j$ orbitals~\cite{heyde2011}, 
such as the very low energy of the first $2^+$ 
state in $^{42}$Si \cite{bastin2007}. 
This jeopardizes the conventional shell gaps 
and related shell model explanation of the 
low-lying excitations using a single nucleon 
or a pair of nucleons occupying just a few valence 
single-particle orbitals above the Fermi surface, 
even for semi-magic nuclei.

Variation of shell gaps with respect to proton-neutron 
asymmetry appears according to the underlying 
nuclear force~\cite{otsuka2020}. 
The weakening of shell gaps can support 
deformed shapes in lighter semi-magic nuclei, 
also resulting in shape coexistence~\cite{heyde2011}, 
such as in $^{44}$S \cite{force2010,rodri2011,gonza2011}. 
The presence of the 2.6 $\mu$s $0_2^+$ 
isomer in $^{44}$S, 
lying very close in energy to the $2_1^+$ state, 
can be manifested using the core excitation 
across the $N=28$ shell gap~\cite{force2010}. 
No such low-lying $0_2^+$ isomer is known 
in the neighboring nucleus $^{46}$Ar \cite{garg2023}. 
Both experimentally and theoretically, 
interpretation of those $N=28$ isotones 
below $^{48}$Ca, from $^{46}$Ar to $^{44}$S, 
and to $^{42}$Si, is rather controversial 
and posed as a formidable challenge 
\cite{bastin2007,gonza2011,rodri2011,gade2019}. 
The weakening of the $N=28$ shell closure 
(i.e. closely lying $f_{7/2}$ and $p_{3/2}$ orbitals), 
and the resulting quadrupole correlations 
can reinforce deformation. 
This mechanism is interpreted within 
the quasi-SU(3) scheme \cite{zuker1995}, 
leading to quasi-degeneracy of the 
$2_1^+$ and $0_2^+$ states, 
which can be also viewed as a nuclear 
Jahn-Teller effect \cite{utsuno2012}. 
On the other hand, for those 
$N=28$ isotones with the $f_{7/2}$ orbital 
occupied by valence protons, 
such as $^{50}$Ti, $^{52}$Cr, and $^{54}$Fe, 
the yrare $0_2^+$ state is also found, 
but not as an isomer, 
mainly because of the large difference 
between the excitation energies of the 
yrare $0_2^+$ state and the yrast $2_1^+$ state. 
These $N=28$ isotonic nuclei with active 
$f_{7/2}$ orbital are usually adopted to be 
nearly spherical~\cite{nomura1970,hensler1971}, 
supporting the yrast $6^+$ isomers~\cite{maheshwari2022}. 
But a weak second minimum on the 
potential energy surface was also reported 
\cite{mizusaki2001,delaroche2007,CEA}, 
which is most likely responsible for yrare states. 
The similar is true for $^{42,44,46}$Ca isotopes, 
with $\nu f_{7/2}$ being active.

In this article, we address the controversial 
cases of $^{42}$Si, $^{44}$S and $^{46}$Ar using 
the interacting boson model (IBM) that 
incorporates the intruder states and configuration 
mixing. 
To the best of our knowledge, that extension of the 
IBM, referred hereafter as the IBM-CM~\cite{isacker}, 
has not so far been considered in these light nuclei. 
We further compare the results for $^{50}$Ti, $^{52}$Cr, 
and $^{54}$Fe, even-even $N=28$ isotones beyond $^{48}$Ca 
having protons in the $fp$ shells, 
to the $^{42,44,46}$Ca isotopes 
having neutrons in $fp$ shells. 
This study also refers to the questionable 
double magicity of $^{48}$Ca using the bosonic 
description of $N=28$ isotones and compares 
to the fermionic wave functions 
calculated with the nuclear shell model (NSM). 
The detailed shell effects and seniority 
states such as the ones observed in the 
light nuclei in question 
have been considered beyond the reach of the 
conventional IBM framework including the one 
employed here, since it is built on the limited 
model space of low-spin pairs of valence nucleons only. 
The scope of this work is to test 
whether the IBM-CM with the 
parameters fine tuned to the observed data 
reproduces the energy level schemes and 
transition properties of the light mass nuclei 
where seniority-type configurations are supposed 
to play a dominant role. 
This study sheds light on the isomerism from 
the point of view of the IBM-CM, 
which has rarely been applied to that nuclear property.

The paper is organized as follows. 
In Sec.~\ref{sec:formalism} we describe the IBM-CM 
framework and procedure to fix its parameters. 
In Sec.~\ref{sec:results} 
the results of the IBM-CM calculations 
on the energy levels and transition properties 
are discussed in comparison to the expeirmental data, 
and to the NSM calculations. 
Section~\ref{sec:summary} gives a summary and 
concluding remarks.

\section{Formalism\label{sec:formalism}}

The interacting boson model (IBM)~\cite{arima1975} 
has been among the most successful and 
employed nuclear structure models together with  
the nuclear shell model \cite{mayer1955}, 
and the geometric collective model \cite{bohr1975}. 
The IBM is comprised of $s$ and $d$ 
bosons with angular momenta $0^+$ and $2^+$, 
representing the monopole and quadrupole pairs 
of valence nucleons, respectively \cite{arimabook,OAI}. 
The number of bosons, $n=n_s+ n_d$, 
is fixed by the microscopic interpretation 
of active valence nucleons $n=n_\pi+n_\nu$, 
where $n_\pi (n_\nu$) equals 
the number of proton (neutron)
particle or hole pairs counted from 
the nearest closed shell. 
A remarkable feature of the IBM is that 
it furnishes the algebra U(6) with generators 
formed by a $s$ and 
$d_\mu$ ($\mu=0,\pm1,\pm2$) bosons. 
The U(6) algebra is further reduced to 
U(5), SU(3) and SO(6) subalgebras, 
providing the vibrational, rotational 
and $\gamma$-unstable spectra, respectively. 
The following IBM Hamiltonian 
has been shown to be 
adequate for phenomenological descriptions of 
the nuclear low-lying states. 
\begin{eqnarray}
\hat{H} = 
\epsilon \hat{n}_d + a_1 \hat{L} \cdot \hat{L} + a_2 \hat{Q} \cdot \hat{Q} \; ,
\label{eq:ibm}
\end{eqnarray}
where $\epsilon$ stands for single $d$-boson 
energy relative to that of $s$ bosons. 
$\hat n_d = d^\dagger\cdot\tilde{d}$ 
is the $d$-boson number operator,
with $\tilde{d}_\mu=(-1)^\mu d_{-\mu}$, U(5) Casimir operator. 
The second term with the strength $a_1$ 
represents the O(3) Casimir operator, 
with $\hat{L} = \sqrt{10} [d^\dagger \times \tilde{d}]^{(1)}$. 
The third term is the quadrupole-quadrupole 
interaction with the strength $a_2$ inducing 
the quadrupole deformation. 
The corresponding operator reads, 
$\hat{Q} = s^\dagger \tilde{d} + d^\dagger \tilde{s} + \chi [d^\dagger \times \tilde{d}]^{(2)}$, 
where $\chi$ is a parameter that determines 
the prolate ($\chi<0$) or oblate ($\chi>0$) 
deformation, reflecting the structure of collective 
nucleon pairs as well as 
the number of valence nucleons~\cite{arimabook}.

Configuration mixing is incorporated in 
such a way \cite{duval1981} that
several independent (unperturbed) IBM Hamiltonians 
for the $n$, $n+2$, $n+4$, $\ldots$ boson 
systems that are associated with 
the 0p-0h, 2p-2h, 4p-4h, $\ldots$ 
particle-hole excitations, respectively, 
are introduced and are allowed to be 
mixed. 
Here we consider up to two configurations, i.e., 
normal $[n]$ and intruder $[n+2]$ configurations, 
for which we assume for $^{44}$S 
the vibrational U(5) (with $\chi=0$) and prolate deformed 
SU(3) (with $\chi=-1.33$) symmetries, respectively. We further assume 
the normal $[n]$ space of $^{42}$Si to be 
dominated by the $\overline{\rm SU(3)}$ (with $\chi=1.33$) symmetry, 
associated with the oblate deformation 
(see e.g., Ref.~\cite{jolie2001}). The full (IBM-CM) Hamiltonian reads, 
\begin{eqnarray}
\hat{H}'=\hat{H}_{n}+(\hat{H}_{n+2}+\Delta)+\hat{V}_{mix} \; ,
\label{eq:hamiltonian}
\end{eqnarray}
where $\hat{H}_{n}$ and $H_{n+2}$ 
represent unperturbed Hamiltonians for the 
normal (0p-0h) and intruder 
(2p-2h) boson spaces, respectively, 
and each has the form given by Eq.~(\ref{eq:ibm}). 
$\Delta$ represents the energy needed 
to promote one nucleon pair from a 
major oscillator shell to the next.
The last term is the interaction that admixes different 
boson spaces, and is given by
\begin{eqnarray}
\hat{V}_{mix} = \alpha (s^\dagger s^\dagger + ss)^{(0)} + \beta (d^\dagger d^\dagger + \tilde{d} \tilde{d})^{(0)} \; ,
\label{mixibm}
\end{eqnarray}
with $\alpha$ and $\beta$ being 
mixing strength. 
The Hamiltonian (\ref{eq:hamiltonian}) 
is diagonalized in the space $[n] \oplus [n+2]$ 
to obtain excitation energies 
and wave functions. 
The parameters used in the present IBM-CM calculations 
are listed in Table~\ref{tab:ibmcm}. 
The $E2$ operator is written as 
$\hat{T}({E2})=e_{2,n} \hat{Q}_{n}+e_{2,n+2} \hat{Q}_{n+2}$ 
with $e_{2,n}$ ($e_{2,n+2}$) and 
$\hat{Q}_{N}$ ($\hat{Q}_{N+2}$) 
being the boson effective charge and 
the quadrupole operator, respectively, 
for the $[n]$ ($[n+2]$) space. 
The $E0$ operator is given as 
$\hat T(E0)=e^{s}_{0,n}\hat n_s + e^{d}_{0,n}\hat n_d$ 
for the $[n]$ space and the same expression 
for the $[n+2]$ space. 
The fixed effective charges 
$e^{s}_{0,n}=0.7$ $e$fm$^2$ and 
$e^{d}_{0,n}=1.3$ $e$fm$^2$ 
are employed for both configurations 
of the $N=28$ isotones below $Z=20$.

To complement the discussion, 
the NSM calculations for $^{42}$Si, 
$^{44}$S and $^{46}$Ar, $N=28$ isotones 
below $^{48}$Ca are performed 
with the NushellX code~\cite{brown} using 
the SDPF-MU interaction~\cite{utsuno2012}. 
The full $sd$ ($pf$) valence space is used 
for protons (neutrons). 
The center-of-mass correction is taken care of as default. 
The NSM calculations for $^{50}$Ti, $^{52}$Cr, 
and $^{54}$Fe, $N=28$ isotones above $^{48}$Ca, 
and $^{42,44,46}$Ca isotopes are performed 
with two $fp$-space interactions, 
GXPF1A~\cite{honma2002} and KB3G~\cite{poves2001}.

\subsection{Procedure of fitting the IBM-CM parameters}

In Table~\ref{tab:ibmcm}, 
the parameters used in the IBM-CM calculations 
are listed. 
As noted above, to reduce the number of parameters, 
here we assume for the parameter $\chi$ 
the SU(3) and $\overline{\rm{SU(3)}}$ limits, 
$\chi=-1.33$ and $1.33$ for the intruder configuration 
of $^{44}$S, and for the normal configuration for $^{42}$Si, 
respectively. 
In addition, the single $d$-boson energy for 
the $\hat{n}_d$ term, a Casimir operator of nearly 
spherical U(5) symmetry, generally has systematic 
of decreasing as a function of valence nucleon 
numbers \cite{arimabook}. 
Therefore, when $\epsilon$ is small, 
and the coefficients $a_1$ 
(for the $\hat{L} \cdot \hat{L}$ term) and 
$a_2$ (for the $\hat{Q} \cdot \hat{Q}$ term) 
are non-zero, with $\chi$ for the latter 
being $-1.33$ or $1.33$, the resulting nuclear 
structure corresponds to the strongly deformed 
prolate [SU(3))], or  oblate [$\overline{\rm{SU(3)}}$] 
shapes.  

With this in mind, $\epsilon$ is generally determined 
to fit the systematic of the measured $2_1^+$ energy level. 
For example, in $^{42}$Si, $\epsilon$ for both the 
$[n]$ and $[n+2]$ configurations is relatively small, 
i.e., $\epsilon < 0.5$ MeV, 
since the experimental $2_1^+$ energy is about 
0.7 MeV, 
while in $^{44}$S and $^{46}$Ar, $\epsilon$ is large for 
both configurations, $\epsilon > 1$ MeV, 
as the observed $2_1^+$ energy level is 
at about $1.4-1.5$ MeV. 
For $^{44}$S, the measured $0_2^+$ level is 
very close in energy to that of the $2_1^+$ state. 
To account for this nuclear structure, the intruder configuration 
for this nucleus is here considered to be of 
strongly deformed SU(3) nature, 
with $\chi=-1.33$ and $\epsilon=0$ MeV, 
so that the calculated $0^+_2$ energy level 
becomes sufficiently low.

For the other $N=28$ isotones $^{50}$Ti, $^{52}$Cr 
and $^{54}$Fe, the experimental $2_1^+$ 
excitation energy is approximately $1.4-1.5$ MeV, 
so we take the $\epsilon$ values to be 
more or less in that range for all these three nuclei, 
irrespective of whether 
$^{40}$Ca or $^{48}$Ca is chosen as the inert core. 
Also for these nuclei, there hardly appears 
to be any signature of strong deformation, 
hence the parameter $\chi$ is assumed to be zero. 
Only the competition between the moments of inertia 
generated by $\hat{L} \cdot \hat{L}$ and 
$\hat{Q} \cdot \hat{Q}$ terms seems to be responsible 
for reproducing the experimental energy spectra. 
Their large $\epsilon$'s hence represent 
the dominance of the $\hat n_d$ term, which favors 
a spherical shape, over the other interaction terms. 
This is the case for the $N=28$ isotones, 
except for $^{42}$Si in both configurations, 
and $^{44}$S in the $[n+2]$ configuration.

In the $^{42,44,46}$Ca isotopes, 
the fixed parameters $a_2=-0.05$ MeV and $\chi=0$ 
are considered for both configurations, 
except for the non-zero value $\chi=-1.0$ for the $[n+2]$ 
space for $^{44}$Ca, which is so chosen 
in order to achieve an agreement with data. 
This may be related to the suppressed location 
of the $2_1^+$ level in the middle of $f_{7/2}$ orbital, 
that is, $(f_{7/2})^4$ configuration. 
The adopted value for the $a_1$ coefficient 
is also almost similar between $^{42}$Ca and $^{46}$Ca, 
corresponding to the two-particle and two-hole spectra 
of the $(f_{7/2})^2$ configuration. 
In $^{44}$Ca, the situation is a bit different due to the fact 
that it is in the middle of the $f_{7/2}$ shell. 
The energy $\epsilon$ is relatively small in Ca isotopes, 
as compared to the $Z>20$, $N=28$ isotones, 
which are dominated by the $f_{7/2}$ orbital. 
This difference is due to the location 
of the $0_2^+$ levels: In the $Z>20$, $N=28$ isotones, 
the $0_2^+$ states are at higher excitation 
energy ranging from $2.6-3.8$ MeV, 
while in the Ca isotopes, the $0_2^+$ states are found 
at 1.8 MeV in $^{42,44}$Ca and at 2.4 MeV in $^{46}$Ca.

As regards the mixing interaction, we have assumed the same amount of the mixing strengths, $\alpha$ and $\beta$ for a given nucleus, controlling the agreement of second-excited states and overall spectra. 
These mixing strength parameters could 
be different between the $s$ and $d$ boson parts, 
whereas in majority of the 
previous IBM-CM calculations the equal mixing strengths  
have been employed for realistic nuclear structure studies. 
The energy off-set $\Delta$ is fitted to 
mainly reproduce the location of the $0_2^+$ level 
in all the chosen nuclei, but for $^{42}$Si and $^{46}$Ar. 
In the latter cases, experimental data are not 
available for the $0^+_2$ excitation energies, and 
hence $\Delta$ is kept the same as that used in $^{44}$S.

\begin{figure*}[!htb]
\centering
\vspace{0cm}
\begin{minipage}{0.5\textwidth}
    \includegraphics[width=\textwidth,height=0.85\textwidth,trim={2cm 0cm 2cm 1cm}]{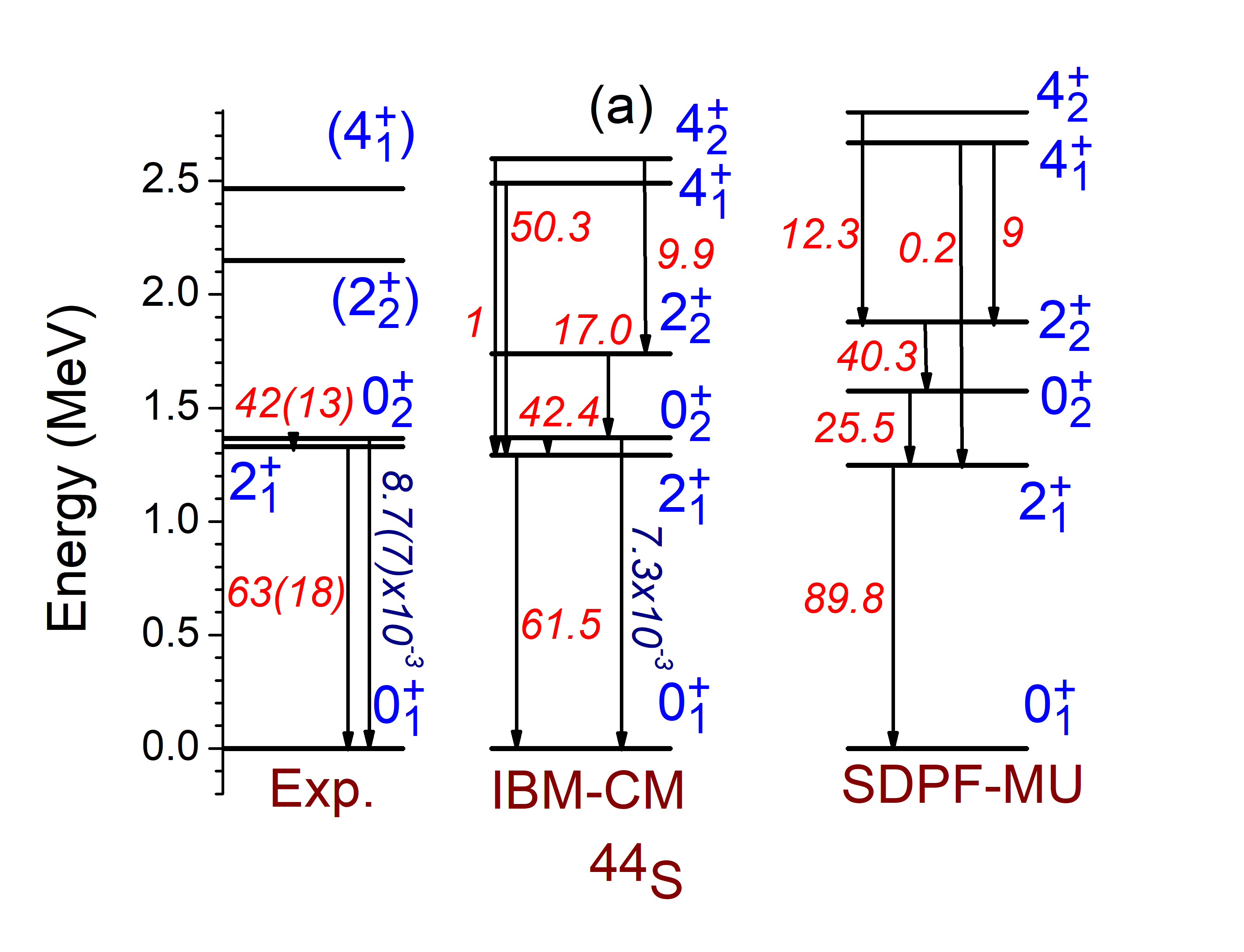}
\end{minipage}
\begin{minipage}{0.5\textwidth}
    \includegraphics[width=\textwidth,height=0.85\textwidth,trim={2cm 0cm 2cm 1cm}]{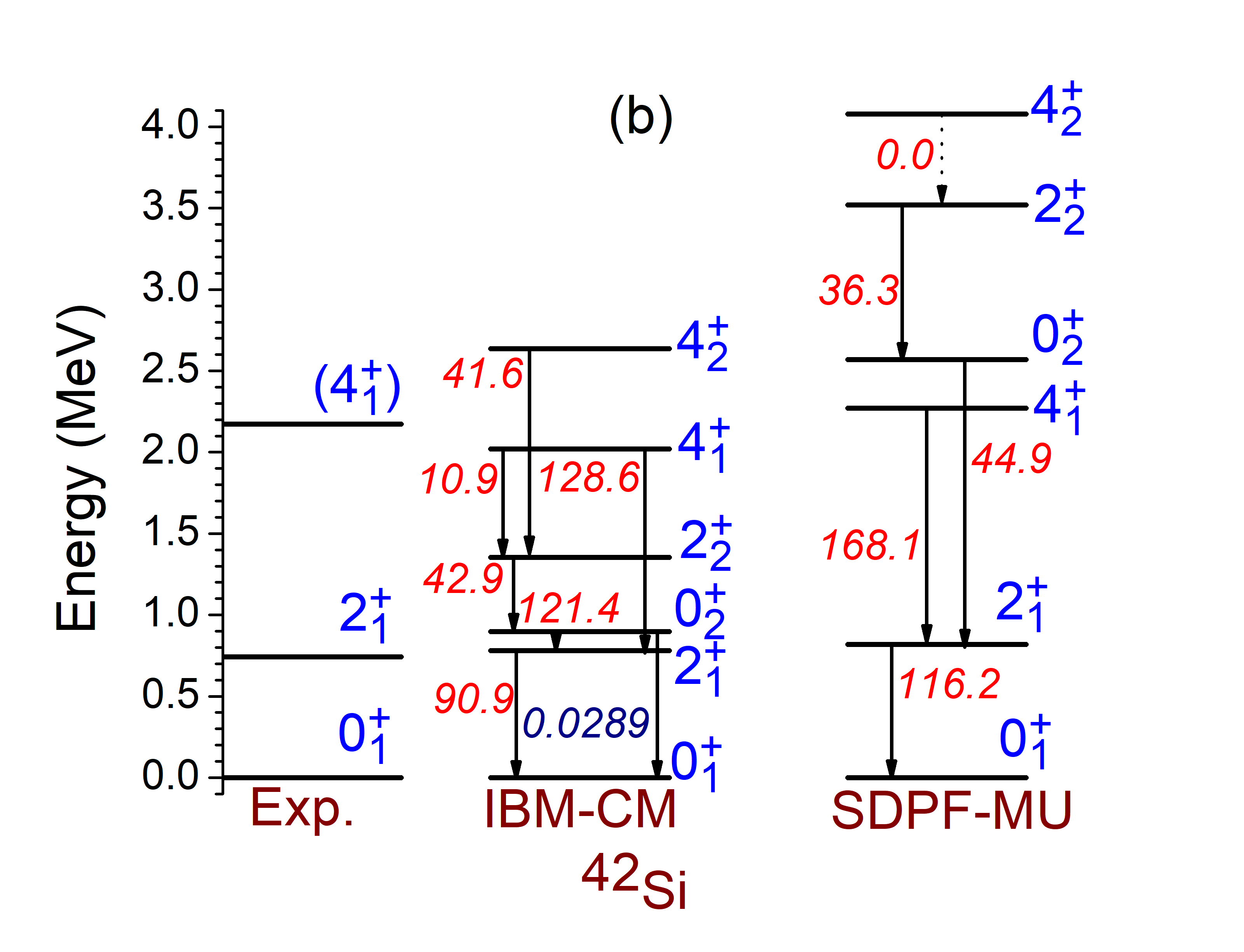}
\end{minipage}
\begin{minipage}{0.5\textwidth}
  \includegraphics[width=\textwidth,height=0.85\textwidth,trim={2cm 1cm 2cm 1cm}]{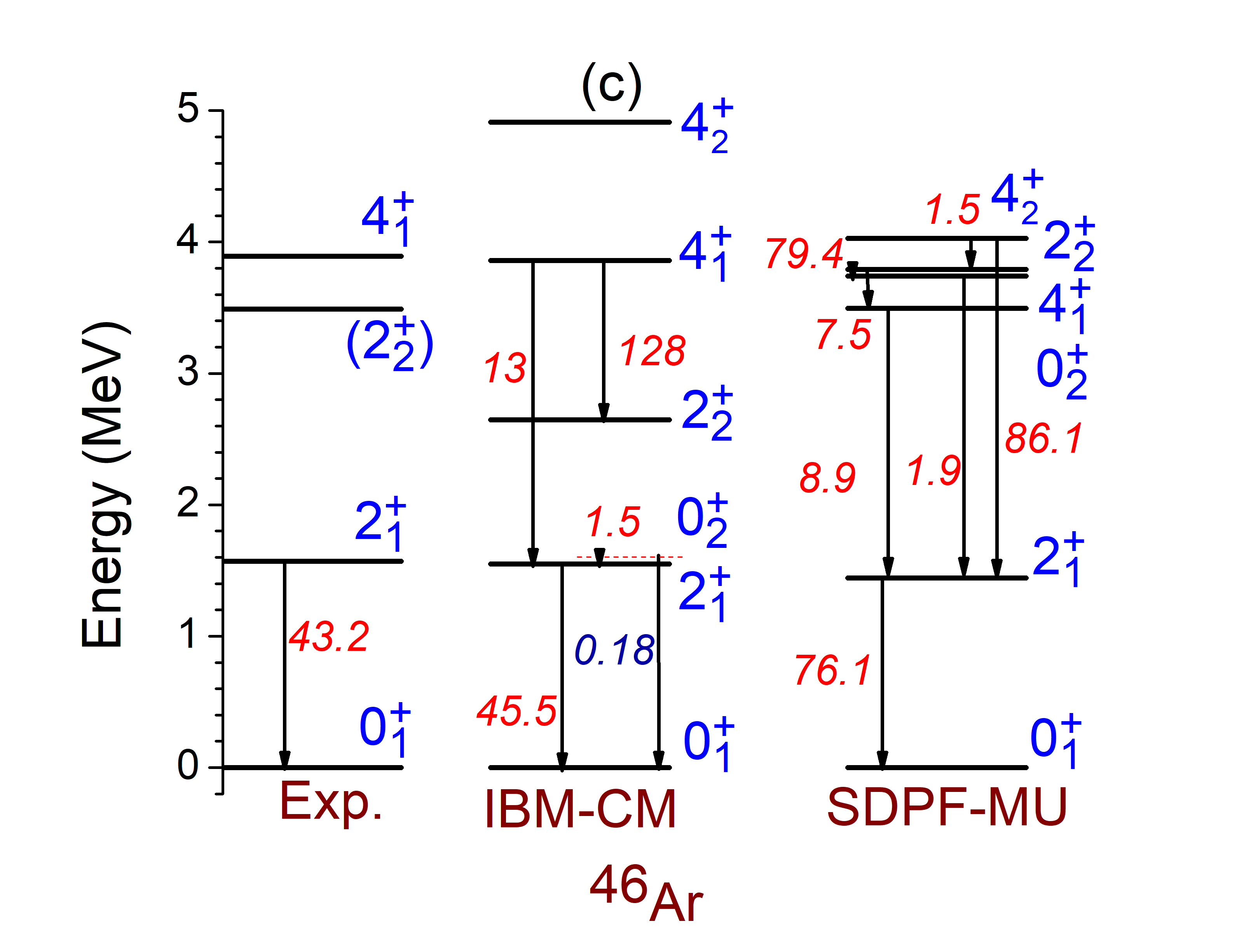}
\end{minipage}
\vspace{0cm}
\caption{\label{fig:ibm}
Comparison of the experimental~\cite{nndc}, calculated level scheme 
for (a) $^{44}$S, (b) $^{42}$Si, and (c) $^{46}$Ar. 
The $B(E2)$ values are shown in the units of 
$e^2$fm$^4$ as red italic numbers along the transition. 
The blue italic number 
denotes the $\rho^2(E0;0_2^+ \rightarrow0_1^+)$ values. NSM results based on SDPF-MU interaction are also shown.} 
\end{figure*}

\section{Results\label{sec:results}}

First, the IBM-CM is applied to the 
$N=28$ isotones $^{42}$Si, $^{44}$S and $^{46}$Ar. 
The doubly-magic nucleus $^{48}$Ca 
is assumed to be the inert core, and hence 
the corresponding boson numbers $N$ for the 
normal configurations of $^{42}$Si, $^{44}$S 
and $^{46}$Ar equal 3, 2, and 1, respectively. 
The predicted and 
experimental level schemes are compared 
in Fig.~\ref{fig:ibm}. 
Since the spectroscopic data for  
$^{42}$Si and $^{46}$Ar are limited, 
let us focus on $^{44}$S as a benchmark case 
for the IBM-CM where $0_2^+$ isomer is known.

\begin{figure*}[!htb]
\centering
\includegraphics[width=\linewidth]{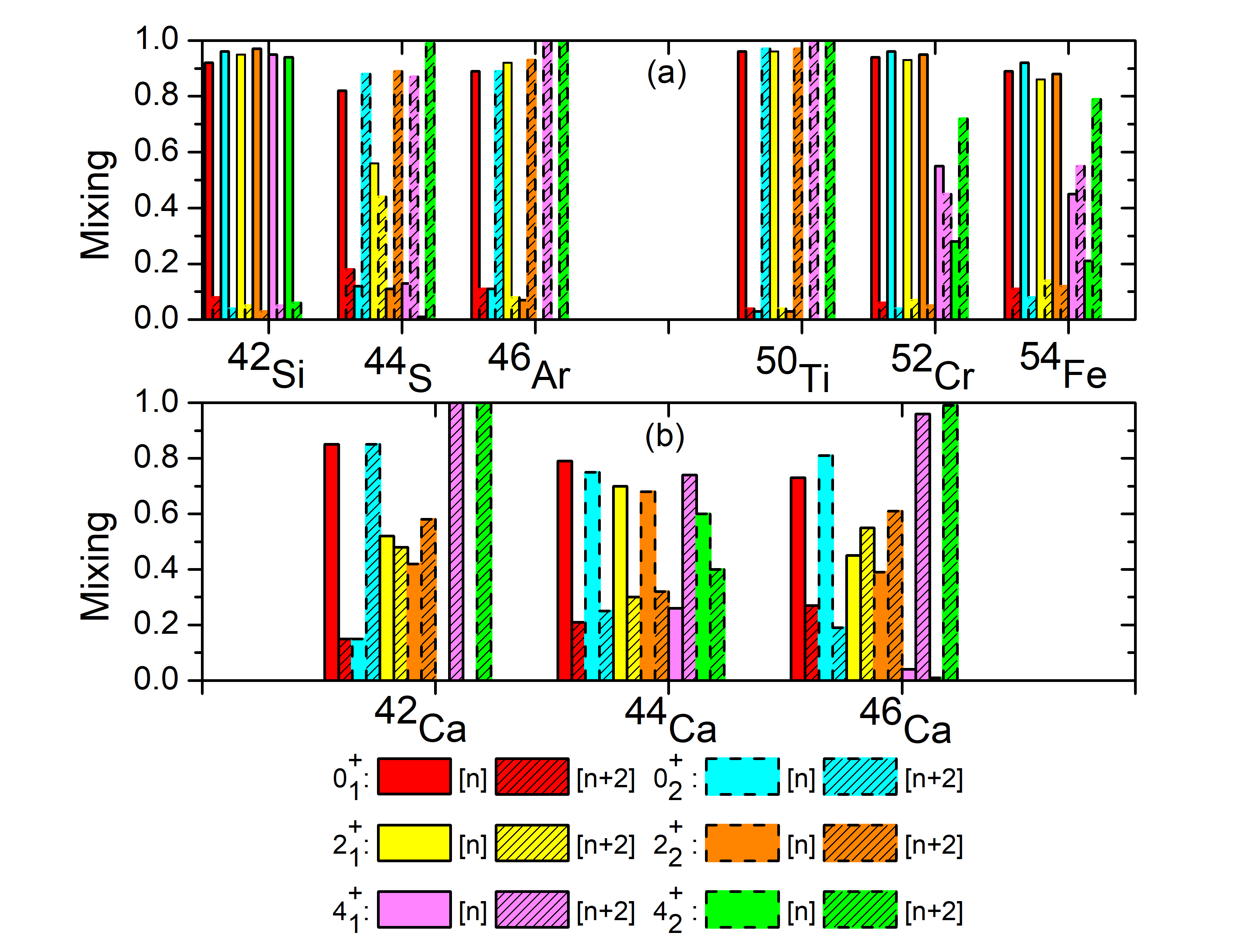}
\caption{\label{fig:mix}
IBM-CM mixing amplitudes arising from both 
the [n] and [n+2] spaces for the low-lying yrast and yrare states in all the chosen nuclei.} 
\end{figure*}

As shown in Fig.~\ref{fig:ibm}(a), 
the present calculation that assumes 
the mixing of the U(5) and SU(3) 
configurations agree nicely with 
the experimental data for the 
low-lying spectra in $^{44}$S. 
For this nucleus 
the optimal values of the mixing parameters 
and the energy offset turn out to be 
$\alpha=\beta=0.2$ MeV, and $\Delta =0.55$ MeV, 
respectively, as listed in Table~\ref{tab:ibmcm}. 
These parameter values are commonly 
used for the $^{42}$Si and $^{46}$Ar. 
Such an amount of mixing between the 
normal and intruder space configurations allows 
for the $0_2^+$ to stay very near to 
the $2_1^+$ state, leading to the isomeric 
nature of the $0_2^+$ state. 
As shown in Fig.~\ref{fig:mix}(a), 
the IBM-CM calculation suggests that 
the $0_1^+$ and $0^+_2$ states are 
mainly dominated by the $[n]$ U(5) and 
$[n+2]$ SU(3) ones, respectively, 
thus indicating the shape coexistence. 
On the other hand, 
both normal and intruder configurations 
are equally mixed in the $2^+_1$ state, 
with the probability of $\approx 50$ \%, 
as depicted in Fig.~\ref{fig:mix}(a). 
The yrast state, $4_1^+$, 
and yrare states, $2_2^+$ and $4_2^+$, 
are again dominated by the deformed ($[n+2]$) 
configuration. 
With the bosonic effective charges, 
$e_{2,n}=e_{2,n+2}=5$ $e$fm$^2$, 
which are fitted to the observed 
$B(E2; 2_1^+ \rightarrow 0_1^+)$ values, 
the measured $B(E2; 0_2^+ \rightarrow 2_1^+)$ rate 
is reproduced reasonably. 
The calculated $\rho^2 (E0)$ values  
for the $0^+_2 \rightarrow 0^+_1$ transition 
agree well with the experimental value. 
These results indicate an ability of the IBM-CM 
to describe quantitatively the low-lying states 
of $^{44}$S and, especially, to 
give the implication for 
the $0_2^+$ isomeric nature of this nucleus.

In $^{44}$S, the present calculation suggests 
the $0_1^+$ state to be primarily characterized 
by a spherical U(5) structure. 
This is explained by the parameters for the 
IBM-CM Hamiltonian used in calculations: 
the parameter $\chi=0$, and $\epsilon$ 
is large, but there is no contributions from 
the $\hat{Q} \cdot\hat{Q}$ term, with the strength 
$a_2=0$ MeV for the $[n]$ configuration. 
The $0_2^+$ isomer, on the other hand, 
is predominantly described by a rotor SU(3) structure, 
indicated by $\chi=-1.33$ and $\epsilon=0$. 
This difference in the structure of wave function 
explains why the $E0$ transition between the $0_2^+$ 
and $0_1^+$ states involves a shape change, 
confirming the $0_2^+$ as a shape isomer. 
The $2_1^+$ state for $^{44}$S is determined by a 
strong mixture between the $[n]$, U(5), and $[n+2]$, 
SU(3) configurations. 
Therefore, a strong $E2$ transition connecting 
the $0_2^+$ to $2_1^+$ states also serves as a signature 
of a shape mixing. 
Such a configuration mixing has been obtained in 
previous IBM-CM calculations in heavy nuclei 
such as those in the Pb-Hg region 
\cite{heyde2011,duval1981,nomura2013,nomura2016}, 
in the neutron-rich $N\approx 60$ \cite{sambataro1982,nomura20161} isotopes, 
in the $Z\approx 50$ region \cite{leviatan2018}, 
and in the $N\approx Z \approx 40$ region \cite{nomura2022}, 
where shape coexistence is empirically 
supposed to occur. 
The current analysis suggests that the configuration 
mixing may play a role in lighter mass region as well.

A similar conclusion has been 
drawn for the nuclear structure of $^{44}$S 
in the previous theoretical studies 
using different nuclear models, e.g., 
within the NSM \cite{force2010} 
based on the 
SDPF-U interaction \cite{nowacki2009}, and within 
the symmetry conserving configuration mixing 
(SCCM) \cite{rodri2011}. 
The SCCM study \cite{rodri2011}, in particular,
obtained nearly 
degenerate $0_2^+$ and $2_1^+$ states,  
but a larger $B(E2; 0^+_2 \rightarrow 2_1^+)$ 
than experimentally suggested. 
The present IBM-CM is able to reproduce the 
$B(E2; 0_2^+ \rightarrow 2_1^+)$ transition 
strength that is remarkably close to the experimental 
data, whereas the calculations from other 
theoretical approaches 
\cite{force2010,rodri2011,gonza2011} predicted 
this transition rate to be rather far 
from the measured value. 
The NSM of Ref.~\cite{force2010} 
could not explain the $\rho^2 (E0)$ 
values, while the SCCM calculation was 
successful in obtaining the measured order. 
The present NSM calculation using the SDPF-MU 
interaction predicts the 
$B(E2; 0_2^+ \rightarrow 2_1^+)$ rate 
of 25.5 $e^2$fm$^4$, which underestimates 
the measured value of $42\pm 13$ $e^2$fm$^4$. Effective charges are $e_\pi =1.35$ and $e_\nu=0.35$, as used in~\cite{chevrier2014}. 
Results for other properties 
obtained from the NSM calculation 
are also compared in Fig.~\ref{fig:ibm}.

\begin{table*}[!htb]
\caption{\label{tab:ibmcm}
Parameters used for the present IBM-CM calculations. 
For the parameters $\epsilon$, $a_1$, 
$a_2$, and $\chi$ those values 
corresponding to the normal $[n]$ 
and intruder $[n+2]$ spaces are shown.}
\centering
\resizebox{0.95\textwidth}{!}{
\begin{tabular}{cccccccccccc}
\hline\hline
\multirow{2}{*}{Nucleus} & 
\multirow{2}{*}{Core} &
\multicolumn{2}{c}{$\epsilon$ (MeV)} &
\multicolumn{2}{c}{$a_1$ (MeV)} &
\multicolumn{2}{c}{$a_2$ (MeV)} &
\multicolumn{2}{c}{$\chi$} &
\multirow{2}{*}{$\alpha=\beta$ (MeV)} & 
\multirow{2}{*}{$\Delta$ (MeV)} \\
\cline{3-4}
\cline{5-6}
\cline{7-8}
\cline{9-10}
 & & 
$[n]$ & $[n+2]$ &
$[n]$ & $[n+2]$ &
$[n]$ & $[n+2]$ &
$[n]$ & $[n+2]$ &
 & \\
\hline
$^{42}$Si & $^{48}$Ca & 0.4 &  0.5 & 0.055 &  0.0 & -0.02 &  +0.05 & 1.33 &  0.0 & 0.2 & 0.55 \\
$^{44}$S & $^{48}$Ca & 1.2 &  0.0 & 0.045 &  0.055 & 0.0 &  0.006 & 0.0 &  -1.33 & 0.2 & 0.55 \\
$^{46}$Ar & $^{48}$Ca & 1.2 &  1.1 & 0.045 &  0.015 & 0.0 &  0.06 & 0.0 &  0.0 & 0.2 & 0.55 \\
\hline
$^{50}$Ti & $^{48}$Ca & 0.90 &  0.75 & 0.08 &  -0.075 & -0.06 &  0.0 & 0.0 &  0.0 & 0.3 & 1.7 \\  
$^{52}$Cr & $^{48}$Ca & 0.90 &  0.75 & 0.08 &  -0.075 & -0.06 &  0.02 & 0.0 &  0.0 & 0.3 & 1.4 \\
$^{54}$Fe & $^{48}$Ca & 0.90 &  0.75 & 0.08 &  -0.075 & -0.06 &  -0.01 & 0.0 &  0.0 & 0.3 & 1.4 \\
\hline
$^{50}$Ti & $^{40}$Ca & 1.65 & 1.45 & -0.06 & 0.08  & 0.01 & 0.06 & 0.0 & 0.0 & 0.3 & 0.9 \\  
$^{52}$Cr & $^{40}$Ca & 1.30 & 0.9 & -0.03 & 0.01 & -0.01 & 0.055 & 0.0 & 0.0 & 0.3 & 0.9 \\
$^{54}$Fe & $^{40}$Ca & 1.20 & 0.90 & 0.048 & -0.085  &  -0.018 & 0.005 & 0.0 & 0.0 & 0.3 & 0.9 \\
\hline
$^{42}$Ca & $^{40}$Ca & 0.40 &  0.40 & 0.20 &  -0.01 & -0.05 &  -0.05 & 0.0 &  0.0 & 0.25 & 1.0 \\
$^{44}$Ca & $^{40}$Ca & 0.60 &  0.40 & 0.05 &  -0.004 & -0.05 &  -0.05 & 0.0 &  -1.0 & 0.25 & 1.1 \\
$^{46}$Ca & $^{40}$Ca & 0.80 &  0.80 & 0.18 &  -0.04 & -0.05 &  -0.05 & 0.0 &  0.0 & 0.25 & 1.0 \\
\hline\hline
\end{tabular}}
\end{table*}

\begin{table*}[!htb]
\caption{\label{tab:sm1} Average shell model proton and neutron-orbital occupancies for $^{42}$Si,$^{44}$S, and $^{46}$Ar using SDPF-MU interaction.}
\centering
\resizebox{0.75\textwidth}{!}{
\begin{tabular}{ c c c c c | c c c c}
\hline
Nucleus & J$^\pi$ & 0d$_{5/2}$ & 0d$_{5/2}$ & 1s$_{1/2}$ & 0f$_{7/2}$ & 1p$_{3/2}$ & 0f$_{5/2}$ & 1p$_{1/2}$ \\
\cline{3-5}
\cline{6-9}
 & &   \multicolumn{3}{c|}{Protons} & \multicolumn{4}{c}{Neutrons}\\
\hline
$^{42}$Si & $0_1^+$ & 4.598 & 0.684 & 0.717 & 4.937 & 0.588 & 2.115 & 0.360 \\
& $0_2^+$ & 4.931 & 0.613 & 0.455 & 5.854 & 0.447 & 1.371 & 0.328 \\
& $2_1^+$ & 4.474 & 0.726 & 0.799 & 4.750 & 0.606 & 2.245 & 0.399 \\
& $2_2^+$ & 4.605 & 0.854 & 0.540 & 5.383 & 0.367 & 1.756 & 0.493 \\
& $4_1^+$ & 4.444 & 0.763 & 0.792 & 4.779 & 0.606 & 2.143 & 0.470 \\
& $4_2^+$ & 4.960 & 0.570 & 0.469 & 5.764 & 0.404 & 1.511 & 0.320 \\
\hline
$^{44}$S & $0_1^+$ & 5.487 & 1.639 & 0.874 & 6.078 & 0.331 & 1.384 & 0.207 \\
& $0_2^+$ & 5.547 & 1.662 & 0.790 & 5.889 & 0.271 &    1.290 &   0.549 \\
& $2_1^+$ & 5.480 &    1.612&   0.907 &     5.746 &   0.341&   1.598 &   0.315  \\
& $2_2^+$ &  5.398 &    1.658 &   0.943 &      5.687 &   0.321 &    1.716 &   0.275\\
& $4_1^+$ &  5.519 &    1.664 &   0.817 &      6.282 &    0.298 &    1.294 &    0.127 \\
& $4_2^+$	&5.567 &    1.532 &    0.899 &      5.450 &    0.344 &   1.699 &    0.506\\
\hline
$^{46}$Ar & $0_1^+$ & 5.853&  2.970 &   1.177 &     6.996 &  0.136 &   0.830 &   0.039 \\
& $0_2^+$ &5.888 &   3.010 &   1.102&    6.341 &   0.146 &   1.372 &   0.142\\
& $2_1^+$ &  5.821&  2.851&   1.327&     6.669&   0.146 &   1.125 &   0.061\\
& $2_2^+$ &  5.782&   2.741&   1.477 &     6.199&   0.184 &   1.529 &   0.088\\
& $4_1^+$ &  5.802&   2.742&   1.455&     6.416&   0.186&   1.357&   0.041\\
& $4_2^+$ &	5.771&  2.808 &   1.421 &    6.123 &   0.169&   1.412 &   0.296 \\
\hline
\end{tabular}}
\end{table*}

In Fig.~\ref{fig:ibm}(a), for $^{44}$S 
the calculated excitation energy of the $4_1^+$ 
state, which decays mainly to the $2_1^+$ state, 
is in agreement with the observed $4^+_1$ energy level, 
with the spin and parity having not 
been firmly established. 
Furthermore, the IBM-CM predicts the energy ratio, 
$R_{4/2}=E(4^+_1)/E(2^+_1)$, 
to be $R_{4/2}<2$ consistently with experiment. 
Other theoretical calculations \cite{rodri2011,gonza2011}, 
as well as the present NSM using the 
SDPF-MU interaction (see Fig.~\ref{fig:ibm}), 
could not reproduce this ratio. 
Within the IBM-CM, the $4_1^+$ state 
is dominated by the $[n+2]$ configuration, 
while in the $2_1^+$ state two configurations are 
strongly mixed. 
This partly explains the measured energy ratio $R_{4/2} < 2$, 
and suggests the importance of the configuration mixing 
in the IBM framework, since with the usual 
$sd$-IBM space consisting 
of a single configuration, the calculated ratio 
$R_{4/2}\geq 2$.

Measurement of the $B(E2;4_1^+ \rightarrow 2_1^+)$ 
transition in $^{44}$S would shed more light
upon this result. 
It is worth noting that the current analysis provides evidence for two $4^+$ states in $^{44}$S separated by an energy gap of 100 keV. The $B(E2; 4_1^+ \rightarrow 2_1^+)$ is about 50 $e^2$fm$^4$, 
whereas the $B(E2; 4_2^+ \rightarrow 2_1^+)$ 
and $B(E2; 4_2^+ \rightarrow 2_2^+)$ 
are only 1 $e^2$fm$^4$ and 10 $e^2$fm$^4$, respectively. 
This observation conforms to the hypothesis of a low-lying $4^+$ isomeric state in $^{44}$S~\cite{parker2017}. In the present calculation, the hindrance observed in the transition from the $4_2^+$ state to $2_1^+$ state can be comprehended by considering different bosonic spaces that are associated with different intrinsic shapes, 
see Fig.~\ref{fig:mix}(a). The calculated $2_2^+$ energy level 
looks a bit more suppressed than the tentatively 
assigned experimental one. Additionally, the $4_2^+$ and $2_2^+$ states are dominated by the $[n+2]$ configuration. Although the $4_2^+$ state could potentially be a longer-lived isomer, it does not qualify as a shape isomer.
The IBM-CM suggests that for $^{44}$S 
the 0p-0h and 2p-2h configurations are strongly 
mixed in the $2_1^+$ state, while other states 
are relatively weakly mixed, that is, 
they are accounted for by a pure normal or 
intruder configuration. 
This supports the weakening of $N=28$ shell 
gap and broadening in nuclear wave functions beyond $f_{7/2}$ 
in the current shell model analysis 
as listed in Table~\ref{tab:sm1}, 
although the amount of core excitations 
can only be judged with full 
$sdpf$-NSM, which is beyond the scope of current study.

The low-lying level structure of $^{42}$Si has been 
extensively investigated by experiments using new-generation 
radioactive-ion beams, such as the one 
at NSCL in 2019 \cite{gade2019}. 
In that reference, the two shell model interactions 
for the $sdpf$ space, 
SDPF-MU~\cite{utsuno2012} and SDPF-U-Si 
(SDPF-U valid for $Z \le 14$) \cite{nowacki2009}, 
could not conclude about the nature of  
the low-lying $0^+$ excited states, especially the $0_2^+$ 
state due to significantly different wave functions.  
The present IBM-CM calculation 
reproduces the low-lying yrast 
$2_1^+$ and $4_1^+$ states using the dominating 
(bosonic) $\overline{\rm SU(3)}$ symmetry corresponding to oblate shape, 
which is assumed for the normal $[n]$ space. 
The $4_1^+$ level has been tentatively 
assigned at 2.1 MeV experimentally. 
The calculated yrast and yrare states shown 
in Fig.~\ref{fig:ibm}(b) are dominated by the 
normal $[n]$ configuration in the IBM-CM.

Since the experimental yrast level scheme for $^{42}$Si 
consisting of the $2^+_1$ and $4^+_1$ states looks like 
rotational band in Fig.~\ref{fig:ibm}(b), 
the IBM with single $\overline{\rm SU(3)}$ 
configuration may appear 
to give a sufficiently good description for $^{42}$Si, 
even without introducing the core excitation. 
Nevertheless, we perform the IBM-CM 
calculation, especially to identify the nature 
of the excited $0^+$ and other yrare states. 
The bosonic effective 
charges $e_{2,n}=e_{2,n+2}=5$ $e$fm$^2$ are used as in $^{44}$S because the measured $B(E2; 2_1^+ \rightarrow 0_1^+)$ in $^{42}$Si is not yet known. 
The effect of including 
the configuration mixing in this case 
is such that, the $0_2^+$ state appears very close in 
energy to the $2_1^+$ state. 
Similar results have been 
obtained from the NSM calculation 
with the SDPF-U-Si interaction in Ref.~\cite{gade2019}. 
No experimental $B(E2)$ has been 
available for $^{42}$Si, thus 
the mystery of the $0_2^+$ state deserves 
further investigation by experiments 
using RI beams \cite{crawford2022}. 
The inclusion of the configuration mixing 
in the IBM for $^{42}$Si 
could also be supported by the self-consistent 
mean-field calculations within the 
Hartree-Fock-Bogoliubov (HFB) method that employed 
the Gogny-D1S energy density functional \cite{delaroche2007}. 
The triaxial quadrupole potential energy surface 
computed by the constrained Gogny-HFB calculation 
\cite{CEA} 
suggested an oblate global minimum at $\beta\approx 0.35$ 
and a prolate secondary minimum with almost the 
same deformation. 
Furthermore more recent {\it ab~initio} valence-space 
in-medium similarity renormalization group 
calculations \cite{yuan2024} have provided a positive 
quadrupole moment $Q(2^+_1)$ for $^{42}$Si, 
suggesting an oblate deformation. 
To conclude, in the IBM-CM for $^{42}$Si, 
only the oblate $\overline{\rm SU(3)}$ symmetry 
is assumed for the $[n]$ space with $\chi=1.33$, 
while the $[n+2]$ space with $\chi=0$ and $a_1=0$ MeV 
indicates a nearly spherical symmetry. 
Configuration-mixing calculations have 
predicted a $0_2^+$ isomer, considering the 
limited experimental data available, 
although this is subject to further measurements. 
The current analysis does not classify this as a shape isomer. 

\begin{figure}[!htb]
\centering
\vspace{-0.2cm}
\includegraphics[width=0.85\textwidth]{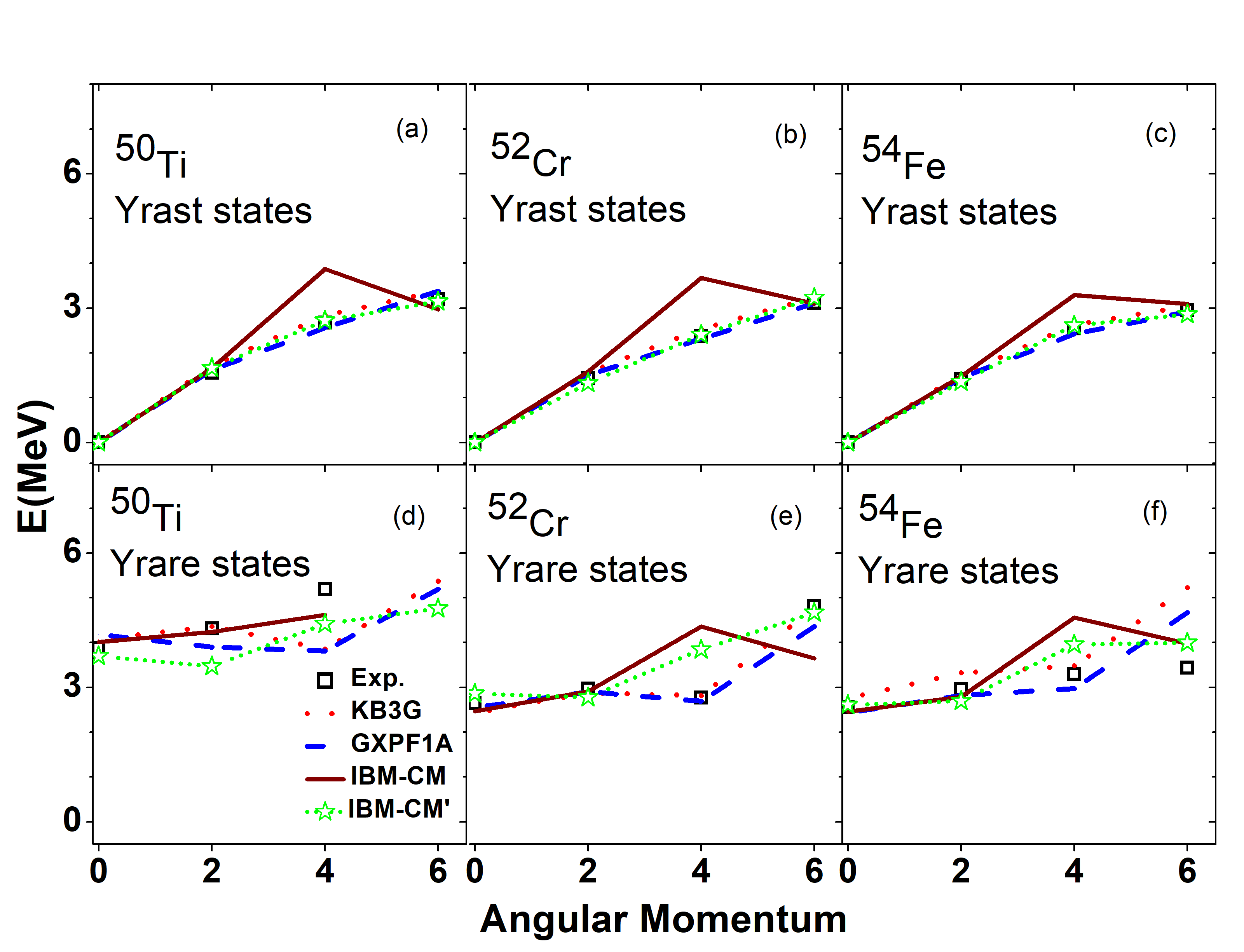}
\vspace{-0.3cm}
\caption{\label{fig:n28paper}
Comparison between the experimental~\cite{nndc} and calculated 
excitation energies for the $N=28$ isotones. 
The NSM results using the 
KB3G and GXPF1A interactions are also shown.} 
\end{figure}


\begin{figure}[!htb]
\centering
\vspace{-0.2cm}
\includegraphics[width=0.85\textwidth]{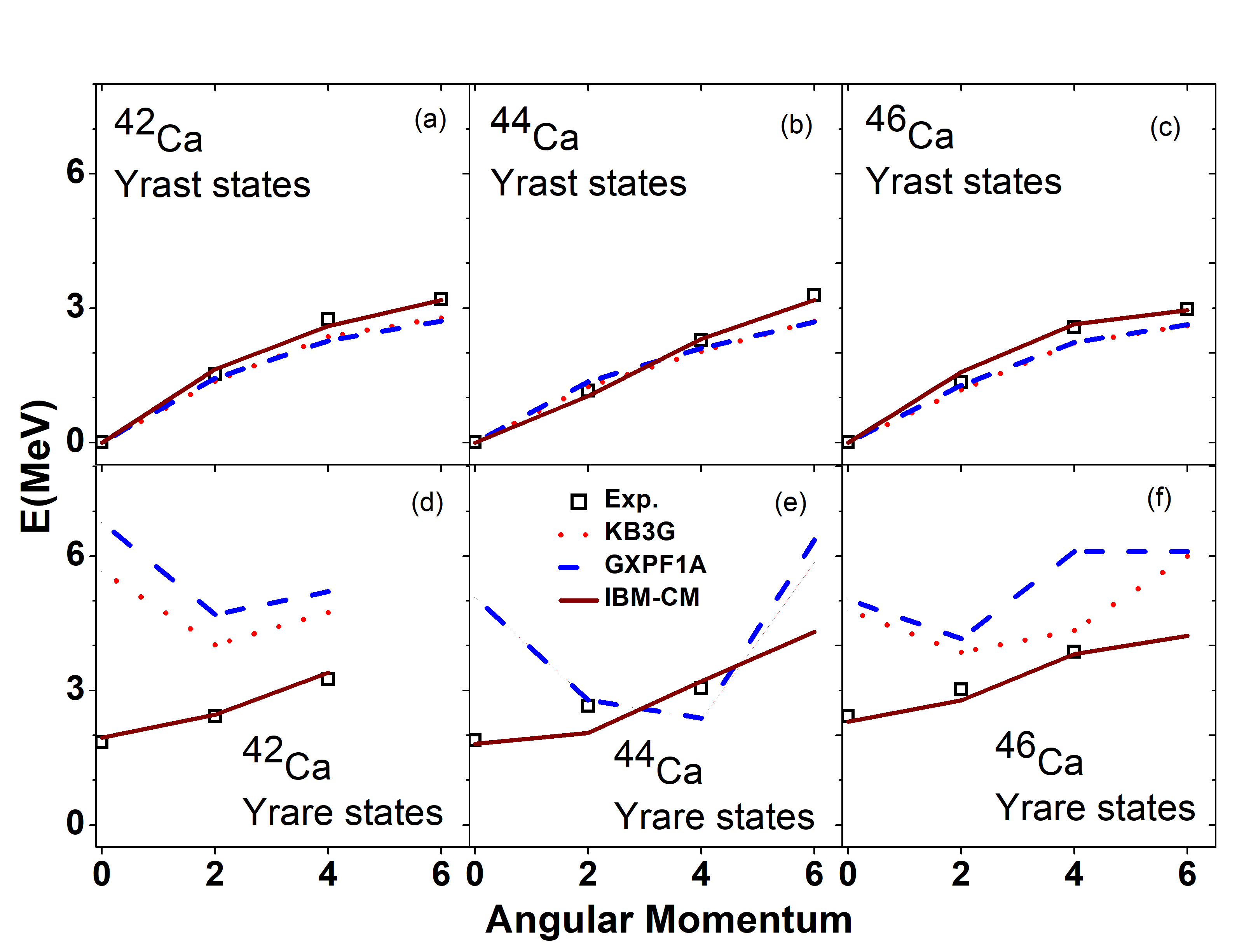}
\vspace{-0.3cm}
\caption{\label{fig:capaper}
Same as Fig.~\ref{fig:n28paper} but for the $^{42,44,46}$ Ca isotopes.} 
\end{figure}

As for the $^{46}$Ar nucleus, 
the calculated energy levels for the 
$2_1^+$ and $0_2^+$ states in the IBM-CM 
are found to be quite degenerate. 
One can see in Fig.~\ref{fig:ibm}(c) that 
the agreement between calculated energies 
and experimental data is reasonable. 
However, the yrast states are expected to 
be of spherical in nature, 
as this nucleus is just one boson away from 
the doubly magic nucleus $^{48}$Ca. 
The yrast states, $0_1^+$ and $2_1^+$, in the 
IBM-CM are mainly dominated by the normal $[n]$ space, 
while the $4_1^+$, $0_2^+$, $2_2^+$, and $4_2^+$ 
states by the intruder $[n+2]$ space. 
Only the $B(E2;2_1^+ \rightarrow 0_1^+)$ 
is experimentally known, which is used 
to fit the bosonic effective charges as 
$e_{2,n}=e_{2,n+2}=6$ $e$fm$^2$. 
The IBM-CM calculation suggests vanishing 
$B(E2;4_2^+ \rightarrow 2_2^+)$  
and $B(E2;2_2^+ \rightarrow 0_2^+)$ values, 
but a substantially large $B(E2; 4_1^+ \rightarrow 2_2^+)$ rate. 
It would be of interest to reveal its (non)isomeric nature 
of the $0_2^+$ state experimentally. 
The calculated $2_2^+$ energy level 
is a bit more suppressed than that of the 
proposed $2_2^+$ state in measured spectra. 
However, there is an energy level at 2.7 MeV 
in experiment with unidentified spin and parity, 
which is very close in energy to the $2_2^+$ state 
predicted by the IBM-CM. 
This may encourage further 
experimental investigation. 
In summary, for $^{46}$Ar, the chosen parameters, 
characterized by the large $\epsilon$ values 
in both the $[n]$ and $[n+2]$ spaces, 
suggest a spherical structure, as compared to 
$^{44}$S and $^{42}$Si. 
Nevertheless, the IBM-CM calculations also suggest a $0_2^+$ isomer, but it would not be a shape isomer. The transition from $0_2^+$ to $0_1^+$ would be hindered by the change in the dominant space and resulting nucleonic configurations.

Figure~\ref{fig:n28paper}(a)-(f) compares the 
experimental and calculated energy spectra 
of both the yrast and yrare $0^+$, $2^+$, $4^+$, 
and $6^+$ states for the $N=28$ isotones, 
$^{50}$Ti, $^{52}$Cr, and $^{54}$Fe. 
The IBM-CM calculation is performed 
with the inert core $^{48}$Ca. 
The agreement between the IBM-CM and experimental 
energies is reasonable for the yrast states, 
except for $4_1^+$ states. 
It is due to the limited number of bosons, 
and could be remedied if one takes 
the $^{40}$Ca doubly-magic nucleus as the inert core 
for these isotones. 
The IBM-CM description of the yrare states, 
$0_2^+$ and $2_2^+$, is also reasonable. 
These isotones are assumed to be nearly spherical in the 
present calculation. 
However, core excitations would make a non-negligible 
contribution to the low-lying states. 
Note that the Gogny-HFB calculation for 
the studied $N=28$ isotones exhibit a flat-bottomed 
potential with a spherical minimum \cite{delaroche2007}, 
in particular, for the nucleus $^{52}$Cr. 
Figure~\ref{fig:n28paper} further depicts the 
NSM results on the excitation energies 
using the KB3G and GXPF1A interactions. 
The NSM explains both the yrast and yrare states 
of these isotonic nuclei well, and also supports 
the spread in neutron wave functions beyond the $f_{7/2}$ shell, 
especially for the yrare states as listed in Table~\ref{tab:sm2}. 
This conclusion is compatible with 
the IBM-CM interpretation.  

In $N=28$ isotones with $Z>20$, a nearly spherical structure is assumed in IBM-CM. Given the assumption of a $^{48}$Ca core with the 
frozen neutron $f_{7/2}$ orbitals, 
calculations using a limited number of bosons 
fail to accurately reproduce the measured $4^+$ state 
with the simple IBM-1 Hamiltonian and the chosen 
interaction strengths. 
Comparative NSM analysis with $fp$-shell interactions 
also showed that the low-lying levels 
did not support the average occupancies of 
neutron $f_{7/2}$ as frozen listed in Table~\ref{tab:sm2}. 
Consequently, further analysis was conducted 
by assuming $^{40}$Ca as an IBM-CM core, 
denoted as IBM-CM$'$ in Fig.~\ref{fig:n28paper}, 
which better explain the spectra for these nuclei, 
including the $4^+$ states. 
The results with a larger number of bosons 
with $^{40}$Ca core suggest that the low-lying 
states in these $N=28$ isotonic nuclei require to 
include the neutron $f_{7/2}$ as an active orbital.

A similar comparison is made for the low-lying yrast 
and yrare state excitations of $^{42,44,46}$Ca 
isotopes in Figs.~\ref{fig:capaper}(a)-(f). 
In this case, $^{40}$Ca is adopted as the inert 
core for the IBM-CM. 
The IBM-CM reproduces the measured 
energy spectra of both the yrast 
and yrare states quite well. 
The NSM with the two $fp$-shell interactions, 
KB3G and GXPF1A, also explain the yrast states, 
but stay very far from the experimental data for 
yrare states. The average occupancies of active neutron shell model orbitals are listed in Table~\ref{tab:sm3}.
The IBM-CM results clearly signify the importance 
of core excitations, or intruder states, 
for the description of yrare states. 
However, the NSM calculation with the $fp$-shell 
interactions could not directly support this finding. 
This is due to the use of $^{40}$Ca as the inert core 
and, to correctly describe the measured systematic 
of those second-excited yrare bands, 
particle-hole excitations from the $sd$ to $pf$ shells 
would have to be taken into account. 
Such an extension would be, however, computationally 
highly demanding \cite{caurier2007}.   

\begin{table*}[!htb]
\caption{\label{tab:sm2} Average shell 
 model proton and neutron-orbital occupancies for $^{50}$Ti, $^{52}$Cr and $^{54}$Fe, N=28 isotonic nuclei from GXPF1A and KB3G interactions.}
\centering
\resizebox{0.75\textwidth}{!}{
\begin{tabular}{c c c c c c c | c c c c}
\hline
Nucleus & J$^\pi$ & Interaction &  0f$_{7/2}$ & 1p$_{3/2}$ & 0f$_{5/2}$ & 1p$_{1/2}$ & 0f$_{7/2}$ & 1p$_{3/2}$ & 0f$_{5/2}$ & 1p$_{1/2}$ \\
\cline{4-11}
 &   & & \multicolumn{4}{c|}{Protons} & \multicolumn{4}{c}{Neutrons}\\
\hline
$^{50}$Ti & $0_1^+$ & GXPF1A & 1.915 &	0.049 &	0.031	& 0.006	&7.573	& 0.238&	0.139&	0.050 \\
&& KB3G & 1.829	& 0.044 &	0.112 &	0.015 &	7.499 & 0.219 &	0.227	& 0.055	\\
&$0_2^+$ & GXPF1A &	1.839&	0.120&	0.025&	0.016&	6.031&	1.496&	0.256&	0.217 \\	
&&KB3G &	1.734&	0.166	&0.075	&0.025	&5.759	&1.572	&0.331 &	0.338\\	
&$2_1^+$& GXPF1A & 1.923 &	0.060&	0.012&	0.004&	7.445&	0.362&	0.138&	0.054 \\
&&KB3G& 1.895&	0.053&	0.042&	0.010&	7.415&	0.304&	0.223&	0.057\\
&$2_2^+$ & GXPF1A & 1.839	& 0.123	& 0.026	&0.013&	6.716&	1.036&	0.154&	0.094\\
&&KB3G& 1.812&	0.085&	0.082&	0.020	&6.566	&1.088&	0.244	&0.102\\
	
&$4_1^+$		&GXPF1A &1.927&	0.046	&0.020	&0.007&	7.542&	0.252&	0.143&	0.063\\
&&KB3G &1.900&	0.042&	0.044	&0.014&	7.481	&0.235&	0.224&	0.060\\

&$4_2^+$		&GXPF1A &1.836&	0.123&	0.028&	0.013&	6.735&	0.982&	0.189&	0.094\\
&&KB3G &1.801&	0.095&	0.085&	0.020&	6.660&	1.002&	0.246&0.091\\

&$6_1^+$		&GXPF1A &1.936&	0.025&	0.035&	0.004&	7.675&	0.171&	0.118&	0.036\\
&&KB3G& 1.914&	0.019&	0.061&	0.006	&7.586&	0.158	&0.213&	0.043\\

&$6_2^+$		&GXPF1A& 1.865&	0.099&	0.026&	0.010&	6.640&	1.085&	0.177&	0.098\\
&&KB3G&	1.868&	0.075&	0.042&	0.015	&6.575&	1.089&	0.236&	0.100\\

\hline

$^{52}$Cr&$0_1^+$		&GXPF1A&	3.747&	0.118&	0.119&	0.016&	7.436&	0.293&	0.214&	0.057\\
&&KB3G&	3.634&	0.093&	0.243&	0.030&	7.373&	0.245&	0.319&	0.063\\

&$0_2^+$		&GXPF1A&	3.376&	0.461&	0.114&	0.049&	5.814&	1.118&	0.649&	0.419\\
&&KB3G& 3.454&	0.334&	0.169&	0.043&	5.707&	1.171&	0.633&	0.488\\

&$2_1^+$	&GXPF1A&	3.706&	0.177&	0.097&	0.019&	7.286&	0.412&	0.232&	0.070\\
&&KB3G&	3.669&	0.130&	0.170&	0.031	&7.264&	0.326&	0.333&	0.077\\

&$2_2^+$	&GXPF1A	&3.530&	0.339&	0.092&	0.039&	6.484&	0.786&	0.462&	0.268\\
&&	KB3G&	3.467&	0.332&	0.157&	0.043&	5.829&	1.069&	0.635&	0.467\\

&$4_1^+$	&GXPF1A&	3.720&	0.153&	0.110&	0.016&	7.356&	0.355&	0.222&	0.067\\
&&KB3G&	3.695&	0.133&	0.146&	0.025&	7.299&	0.303&	0.324&	0.075\\

&$4_2^+$	&GXPF1A&	3.785&	0.144&	0.053&	0.018&	7.233&	0.436&	0.274&	0.057\\
 &&KB3G&	3.767&	0.087&	0.119&	0.026&	7.311&	0.306&	0.325&	0.058\\

&$6_1^+$	&GXPF1A&	3.838&	0.086&	0.067&	0.010&	7.507&	0.259&	0.184&	0.049\\
&&KB3G&	3.780&	0.066&	0.136&	0.018&	7.449&	0.203&	0.294&	0.055\\

&$6_2^+$	&GXPF1A&	3.536&	0.320&	0.108&	0.036&	6.540&	0.896&	0.392&	0.172\\

&&KB3G&	3.583&	0.224&	0.154&	0.040&	6.526&	0.899&	0.406&	0.169\\
\hline

$^{54}$Fe&$0_1^+$	&GXPF1A&	5.621	&0.190	&0.164	&0.026&	7.431&	0.293&	0.225&	0.051\\
&&	KB3G&	5.524&	0.124&	0.313&	0.039&	7.398&	0.199&	0.346&	0.056\\

&$0_2^+$		&GXPF1A&	5.250&	0.493&	0.191&	0.066&	5.797&	1.141&	0.653&	0.409\\
&&KB3G	&5.314	&0.341&	0.285&	0.059&	5.635&	1.243&	0.619&	0.503\\

&$2_1^+$		&GXPF1A&	5.586&	0.253&	0.132&	0.029&	7.313&	0.388&	0.238&	0.060\\
&&KB3G&	5.579&	0.156&	0.226&	0.039&	7.353&	0.242&	0.345&	0.060\\

&$2_2^+$		&GXPF1A&	5.165&	0.533&	0.234&	0.069&	6.421&	0.933&	0.430&	0.217\\
&&KB3G&	5.304&	0.360&	0.276&	0.060&	5.699&	1.155&	0.650&	0.495\\	

&$4_1^+$		&GXPF1A&	5.698&	0.161&	0.117&	0.025&	7.438&	0.288&	0.224&	0.051\\
&&KB3G&	5.652&	0.105&	0.208&	0.035&	7.432&	0.184&	0.330&	0.054\\

&$4_2^+$		&GXPF1A&	5.424&	0.354&	0.177&	0.045&	6.616&	0.857&	0.357&	0.170\\
&&KB3G&	5.430&	0.230&	0.287&	0.053&	6.539&	0.907&	0.379&	0.176\\
&$6_1^+$	&GXPF1A&	5.727&	0.149&	0.107&	0.017&	7.508&	0.251&	0.200&	0.041\\
&&KB3G&	5.673&	0.098&	0.201&	0.028&	7.461&	0.166&	0.325&	0.049\\
	
&$6_2^+$	&GXPF1A&	5.382&	0.392&	0.176&	0.050&	6.514&	0.700&	0.673&	0.113\\
&&KB3G&	5.389&	0.237&	0.321&	0.053&	6.610&	0.394&	0.878&	0.118\\

\hline
\end{tabular}}
\end{table*}

\begin{table*}[!htb]
\caption{\label{tab:sm3} Average shell model neutron-orbital occupancies for $^{42,44,46}$Ca from GXPF1A and KB3G interactions.}
\centering
\centering
\resizebox{0.75\textwidth}{!}{
\begin{tabular}{c c c c c c | c c c c}
\hline
Nucleus & J$^\pi$ & 0f$_{7/2}$ & 1p$_{3/2}$ & 0f$_{5/2}$ & 1p$_{1/2}$ & 0f$_{7/2}$ & 1p$_{3/2}$ & 0f$_{5/2}$ & 1p$_{1/2}$ \\
\cline{3-10}
 &  & \multicolumn{4}{c|}{GXPF1A} & \multicolumn{4}{c}{KB3G}\\
\hline
$^{42}$Ca & $0_1^+$ & 1.945 & 0.031 & 0.019 & 0.005 & 1.846 & 0.062 & 0.074 & 0.018 \\
 & 0$_2^+$ & 0.042 & 1.762 & 0.028 & 0.169 & 0.083 & 1.798 & 0.002 & 0.117 \\
 & 2$_1^+$ & 1.965 & 0.032 & 0.002 & 0.001 & 1.914 & 0.068 & 0.011 & 0.007 \\
 & 2$_2^+$ & 0.989 & 0.976 & 0.014 & 0.021 & 1.032 & 0.940 & 0.013 & 0.015 \\
 & 4$_1^+$ & 1.977 & 0.014 & 0.005 & 0.003 & 1.953 & 0.028 & 0.010 & 0.009 \\
 & 4$_2^+$ & 1.010 & 0.938 & 0.008 & 0.044 & 1.022 & 0.902 & 0.013 & 0.063 \\
 & 6$_1^+$ & 1.991 & 0.000 & 0.009 & 0.000 & 1.985 & 0.000 & 0.015 & 0.000 \\
 & 6$_2^+$ & 1.009 & 0.000 & 0.991 & 0.000 & 1.015 & 0.000 & 0.985 & 0.000 \\
\hline
$^{44}$Ca & $0_1^+$ & 3.889 & 0.057 & 0.046 & 0.009 & 3.756 & 0.090 & 0.129 & 0.026 \\
 & 0$_2^+$ & 2.893 & 1.068 & 0.021 & 0.018 & 2.761 & 1.153 & 0.048 & 0.038 \\
 & 2$_1^+$ & 3.895 & 0.076 & 0.024 & 0.005 & 3.806 & 0.108 & 0.067 & 0.018 \\
 & 2$_2^+$ & 3.895 & 0.083 & 0.010 & 0.011 & 3.833 & 0.115 & 0.028 & 0.025 \\
 & 4$_1^+$ & 3.888 & 0.073 & 0.035 & 0.004 & 3.834 & 0.109 & 0.046 & 0.011 \\
 & 4$_2^+$ & 3.948 & 0.037 & 0.010 & 0.005 & 3.885 & 0.050 & 0.050 & 0.015 \\
 & 6$_1^+$ & 3.948 & 0.025 & 0.024 & 0.003 & 3.892 & 0.037 & 0.061 & 0.010 \\
 & 6$_2^+$ & 2.938 & 0.956 & 0.041 & 0.064 & 2.919 & 0.959 & 0.044 & 0.078 \\
 \hline
$^{46}$Ca & $0_1^+$ & 5.886 & 0.074 & 0.050 & 0.010 & 5.738 & 0.092 & 0.143 & 0.026 \\
 & 0$_2^+$ & 4.906 & 1.041 & 0.032 & 0.021 & 4.870 & 1.026 & 0.069 & 0.035 \\
 & 2$_1^+$ & 5.867 & 0.097 & 0.030 & 0.006 & 5.791 & 0.108 & 0.082 & 0.019 \\
 & 2$_2^+$ & 4.904 & 1.011 & 0.051 & 0.033 & 4.790 & 1.033 & 0.105 & 0.072  \\
 & 4$_1^+$ & 5.912 & 0.052 & 0.028 & 0.008 & 5.839 & 0.067 & 0.074 & 0.020 \\
 & 4$_2^+$ & 4.915 & 0.989 & 0.068 & 0.028 & 4.832 & 1.014 & 0.111 & 0.043 \\
 & 6$_1^+$ & 5.924 & 0.043 & 0.029 & 0.004 & 5.856 & 0.055 & 0.075 & 0.014 \\
 & 6$_2^+$ & 4.929 & 1.004 & 0.049 & 0.018 & 4.876 & 0.980 & 0.081 & 0.063 \\
\hline
\end{tabular}}
\end{table*}

\begin{figure}[!htb]
\centering
\includegraphics[width=0.75\textwidth,trim={2cm 1.5cm 2.5cm 0cm}]{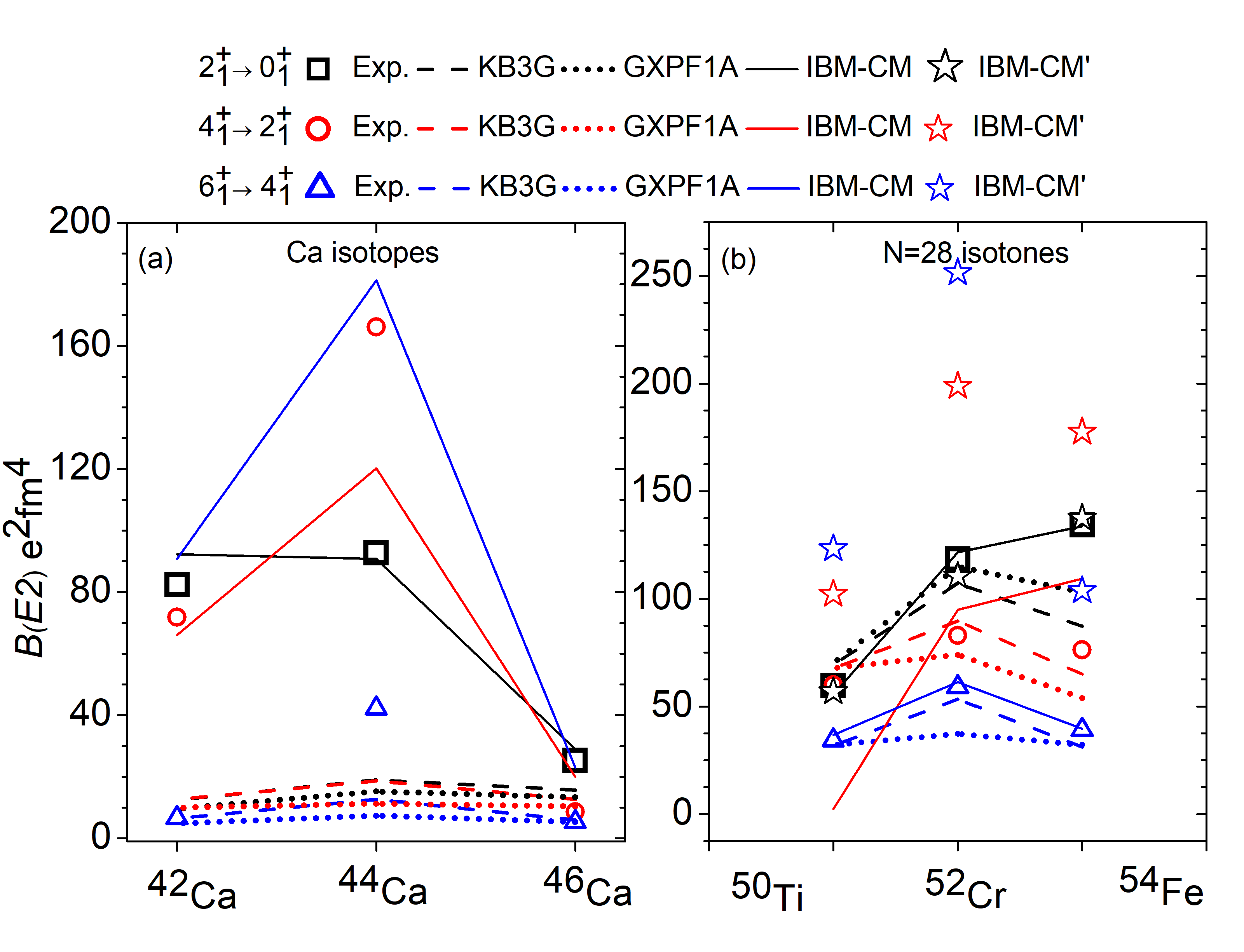}
\caption{\label{fig:be2sm}
Comparison of the experimental~\cite{nndc} and calculated 
$B(E2;I \to I-2)$ values for (a) the 
$^{42,44,46}$Ca isotopes, and for (b) the 
$^{50}$Ti, $^{52}$Cr and $^{54}$Fe, $N=28$ isotones. 
In NSM (based on GXPF1A and KB3G), the effective neutron (proton) 
charge $e_\nu=0.65$ ($e_\pi=1.3$) 
is used for the Ca isotopes ($N=28$ isotones). } 
\end{figure}

Figure \ref{fig:be2sm} compares the 
experimental~\cite{nndc} and calculated 
$B(E2; I \rightarrow I-2)$ 
transition rates for (a) the $^{42,44,46}$Ca isotopes, 
and (b) the $N=28$ isotones $^{50}$Ti, $^{52}$Cr and $^{54}$Fe. 
The bosonic effective charges 
$(e_{2,n}, e_{2,n+2})$ = 
$(10.0,5.5)$, $(5.5,5.5)$, and $(3.5,1.5)$ for $^{42,44,46}$Ca 
isotopes, respectively, and 
$(e_{2,n}, e_{2,n+2})$ = $(7.5,3.5)$, $(7.5,4.5)$, and $(6.5,2.3)$ 
for $^{50}$Ti, $^{52}$Cr and $^{54}$Fe, respectively,
are obtained from the fit to 
the $B(E2; 2_1^+ \rightarrow 0_1^+)$ value 
for each nucleus. 
The different bosonic effective charges between the 
two configurations are considered here, which are 
determined so as to account for the mixing amplitudes of the 
normal $[n]$ and intruder $[n+2]$ bosonic spaces in the 
$0^+_1$ and $2^+_1$ states, as plotted in Fig.~\ref{fig:mix}(a) for $^{50}$Ti, $^{52}$Cr and $^{54}$Fe isotones, and Fig.~\ref{fig:mix}(b) for $^{42,44,46}$Ca isotopes. 
The larger value of $[n]$ space effective 
charge reflects the dominance of normal 
configuration for the states in the given transition. The $B(E2)$ estimates with $^{40}$Ca core, IBM-CM$'$ are also shown with stars in Fig.~\ref{fig:be2sm}(b) using the bosonic effective charges $(e_{2,n}, e_{2,n+2})$ =$(3.8,0.9)$, $(4.7,0.7)$, and $(5.0,1.5)$, for $^{50}$Ti, $^{52}$Cr and $^{54}$Fe, respectively. One can judge the dominance of $[n]$ configuration in IBM-CM$'$ set from the used effective charges, as should be the case with $^{40}$Ca core for $Z>20$, $N=28$ isotones, since we have already included below the core-excitations by shifting the core for these isotones. 
For the Ca isotopes, the predicted 
$B(E2; 4_1^+ \rightarrow 2_1^+)$ transitions 
are close to the measured ones, but the 
$B(E2; 6_1^+ \rightarrow 4_1^+)$ rates 
are considerably overestimated. 
In the $N=28$ isotones, the $B(E2)$ values 
computed with the IBM-CM, in most cases, 
agree with the experimental data. 
A discrepancy is found for the 
$4_1^+ \rightarrow 2_1^+$ transition in $^{50}$Ti. 
This could be corrected by using 
a larger number of bosons. 
The $fp$-shell model results are in reasonable agreement for the yrast level scheme in $Z>20$, $N=28$ isotones; $^{50}$Ti, $^{52}$Cr and $^{54}$Fe. 
Though the yrare states do not support $f_{7/2}$ as fully filled, i.e. $N=28$ subshell closure, as can be followed from the average occupancies listed in Table~\ref{tab:sm2}. 
The IBM-CM analysis provides a consistent result. 

The NSM prediction on the $B(E2)$ values, 
for both the KB3G and GXPF1A effective interactions, 
underestimates the data for the Ca isotopes, 
but is in a reasonable agreement with data 
for the $N=28$ isotones. 
The $6_1^+$ isomeric nature in the isotonic nuclei, 
$^{50}$Ti and $^{54}$Fe, is very well explained 
by the NSM based on the two interactions. 
For $^{52}$Cr, there are two $4^+$ states 
lying close to each other, 
providing an extra decay branch for 
the $6_1^+$ state and no isomer. 

For the $^{42,44,46}$Ca isotopes, 
the IBM-CM results suggest the importance 
of core excitations. This is also evident 
in the $fp$-shell model results, 
based on both the GXPF1A and KB3G interactions, 
which are not sufficient in explaining the transition 
probabilities between the yrast levels 
(cf. Fig.~\ref{fig:be2sm}). 
The incorporation of core excitations in IBM is 
relatively feasible as compared to the NSM, 
even for lighter mass nuclei.

\section{Conclusions\label{sec:summary}}

Structure of $^{42}$Si, $^{44}$S, 
and $^{46}$Ar has been studied in terms 
of the IBM-CM, which has rarely been 
applied in light-mass regions. 
The nature of $^{42}$Si has been attributed 
to be of deformed, SU(3), character. 
The interpretation within the IBM-CM 
that the $0_2^+$ (shape) isomer 
in $^{44}$S arises from the shape coexistence 
provides a crucial piece of knowledge 
about the shape isomers across the nuclear landscape, 
which remains an open issue in 
nuclear structure physics. 
$^{46}$Ar is shown to be nearly spherical, 
but the nature of the $0_2^+$ state deserves 
further investigations by experiment. 
The $fp$-shell shell model calculations 
are in a reasonable agreement for the yrast 
level scheme in $Z>20$, $N=28$ isotones; 
$^{50}$Ti, $^{52}$Cr and $^{54}$Fe. 
Though the yrare states do not support 
$f_{7/2}$ as fully filled, i.e. N=28 subshell closure, 
as can be followed from the average occupancies listed in Table~\ref{tab:sm2}. 
This finding is consistent with that of 
the IBM-CM analysis. 
Structure of the $N=28$ isotones above $^{48}$Ca 
differs significantly from those nuclei 
below $Z=20$, even though both regions are sensitive 
to core excitations. 
The IBM-CM calculation for $^{42,44,46}$Ca isotopes 
sheds light upon the interpretation of 
their yrare states and 
transition probabilities between the yrast states.

This work will pave a way to study nuclear structure 
properties in this mass region which are 
of much interest for future measurements. 
For instance, these results on the shell-breaking mechanism 
at the first spin-orbit closed shell $N=28$ 
are relevant for charge radii 
puzzle around $^{48}$Ca \cite{ruiz2016}. 
In addition, it is also of crucial importance to 
assess the suitability of the IBM-CM consisting of 
the collective $s$ and $d$ bosons for the 
interpretation of the low-lying states of light nuclei, 
especially for those in which 
seniority-type configurations are expected to play 
a dominant role. 
This could be addressed by 
analyzing the IBM-CM wave functions in connection 
to the underlying shell structure, or including 
additional degrees of freedom beyond $s$ and $d$ ones. 
These analyses 
present an interesting future work. 

\section{Acknowledgements}

The author BM gratefully acknowledges 
the financial support from the Croatian Science Foundation 
and the \'Ecole Polytechnique F\'ed\'erale de Lausanne, 
under the project 
TTP-2018-07-3554 ``Exotic Nuclear Structure and Dynamics'', 
with funds of the Croatian-Swiss Research Programme.

\end{document}